\documentclass[journal]{IEEEtran}
\usepackage{cite}

\usepackage{booktabs,bbold}

\ifCLASSINFOpdf
   \usepackage[pdftex]{graphicx}
\else
\fi
\usepackage[cmex10]{amsmath,mathtools}
\usepackage[tight,footnotesize]{subfigure}

\usepackage[caption=false,font=footnotesize]{subfig}

\usepackage{tikz}

\DeclareMathOperator*{\argmax}{arg\,max}

\DeclarePairedDelimiter{\ceil}{\lceil}{\rceil}
\begin{document}
%
\title{Determining Optimal Stop-Loss Thresholds via Bayesian Analysis of Drawdown Distributions*}
%
%
%

\author{Antoine~E.~Zambelli\thanks{*This is a preprint. Please do not invest your life savings based on this.} %
	}

%
%

\markboth{Journal of \LaTeX\ Class Files,~Vol.~6, No.~1, January~2007}%
{Shell \MakeLowercase{\textit{et al.}}: Bare Demo of IEEEtran.cls for Journals}
%



\newcommand{\Emmett}[5]{
	\draw[#4] (0,0)
	\foreach \x in {1,...,#1}
	{   -- ++(#2,rand*#3)
	}
	node[right] {#5};
}

\maketitle
\thispagestyle{empty}

\begin{abstract}
Stop-loss rules are often studied in the financial literature, but the stop-loss levels are seldom constructed systematically. In many papers, and indeed in practice as well, the level of the stops is too often set arbitrarily. Guided by the overarching goal in finance to maximize expected returns given available information, we propose a natural method by which to systematically select the stop-loss threshold by analyzing the distribution of maximum drawdowns. We present results for an hourly trading strategy with two variations on the construction.
\end{abstract}

\begin{IEEEkeywords}
Bayesian, Finance, Hedging, Maximum Drawdown, Stop-Loss.
\end{IEEEkeywords}

%

\section{Introduction}
%
%
%
%
\IEEEPARstart{F}{inancial} traders or quantitative researchers have one overarching goal, to maximize expected returns. Now, this can only be gone given available information at present time. Unfortunately, analyzing every piece of information is unfeasible, and choosing which information to process is the crux of their work. In our case, we choose to look at the Maximum Drawdown and how it can be applied to stop-losses.

Stop-losses are a financial element used to mitigate losses by signaling to a trading strategy that it should exit a position. If the price of a traded asset passes the stop-loss threshold, then we conclude that the position is likely to result in further losses in the future and we exit now (possibly locking in a loss at present). In this way, they are an indirect method of maximizing returns (by losing less when the strategy is wrong).

Note that implementing stops is not always beneficial and may reduce overall expected returns of a strategy~\cite{kam}. They conclude that different regimes can determine whether stops will be useful or not and we will keep this in mind when analyzing our results. We also see ``what-if" studies that look into how stop-loss rules could have helped in certain past events~\cite{han}. In more practitioner-oriented writings, authors focus on forecasting the price path in order to set the stop threshold~\cite{roth}.

However, given that a trading strategy is looking to predict market movements, we feel that setting stop-loss thresholds by again predicting the final price point is in essence doubling-down. In this paper, we will try to find a way to systematically determine the optimal stop-loss threshold for a strategy and its traded asset, such that we do not resort to setting arbitrary stop-loss levels and do not have to resort to complex price predictions.

\subsection{Signal-Only Strategy}

One requirement for the method we will develop is to have a trading strategy with both entry and exit signals. We will use these Signal-Only trades to calibrate our model. While this is unlikely to be restrictive in terms of strategies, it will increase the computational requirements to some degree.

As a case study, and for reproducibility, we will examine the following Signal-Only system using simple moving averages (SMA). Every hour, we compute the 20-hour SMA. If the asset price is above the SMA then we enter a Long position. If at any point in the future (our highest frequency of data is 1-minute) we break below the last computed SMA value, we exit our position. Should we (at an hourly point) break back over the SMA, then we reenter a Long position.

Computationally, we will run the Signal-Only system in parallel and keep track of what trades it would have made - to build the histogram, which we will see below.

\section{Background and Motivation}

\begin{figure}[h!]
\begin{center}
\begin{tikzpicture}
\draw[dotted] (0,0)--(8,0);
\draw[dashed] (0,-0.5)--(8,-0.5);
\draw[solid] (0,-1.6)--(8,-1.6);
\Emmett{400}{0.02}{0.2}{red}{}
\Emmett{400}{0.02}{0.2}{blue}{}
\end{tikzpicture}
\end{center}
\caption{Simulated asset price trajectories from entry to exit.}
\label{fig:sim}
\end{figure}
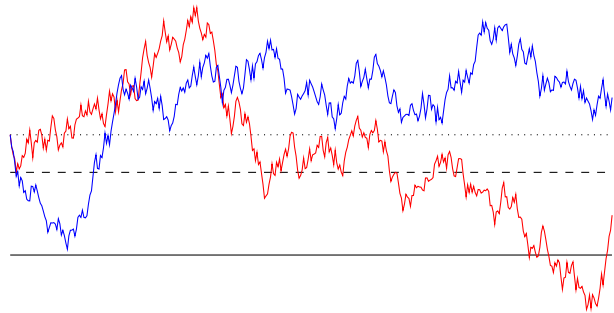

To illustrate some of the issues we face with stop-losses, we can look at the two simulated asset price trajectories in Figure~\ref{fig:sim}. If these trajectories were plotted from trade entry to exit point, then the blue path would be a winner and the red a loser - closing above and below their entry price, respectively.
\pgfmathsetseed{1398} 

However, we can see that this win/loss designation can change over time. The blue path started on a drawdown, and eventually rebounded to end the period above the entry price. In this case, tight stops (say, the dashed line) would have led to stop-loss remorse. In order to capture the win, we would have had to set our stop level at the that of the solid line. Meanwhile, the red path started strong and eventually lost its gains and continued on to close at a loss. We would have needed to exit our position earlier, locking in the early gains before the price started moving against us.

\subsection{Trailing Stops}

As discussed, one issue of interest is locking in gains. A common tool to accomplish this is what is commonly known as trailing stop-losses. Instead of calibrating stop levels at, say, $97\%$ of the asset price at time of entry, we instead update the threshold such that it is always at $97\%$ of the highest asset price seen since entry. In this way, we would lose no more than $3\%$ from the maximum, instead of possibly losing out on all the gains made since we entered.

Note that infinitely tight trailing stops are the opposite limit of the Signal-Only system, which itself has no stops at all. While in theory this would limit our losses at the first sign of a downward move, we would experience severe stop-loss remorse. That is the term used to describe the effect of ``cutting off winners". That is, some positions will have a temporary loss, only to rebound down the line and exit even higher than they were at the start of the drawdown.

\subsection{Categorization}

We have established that we have several types of trajectories: winners and losers. Furthermore, winners can have drawdowns before they exit and losers can have upswings before they crash. The piece of information shared by all these trajectories is that they all have a maximum drawdown at every point in time (theoretically possibly $0$).

We can handle losers' upswings with trailing stops, but we now need to find a way to tackle the problem of winners' drawdowns. We cannot set very tight stops, but we likely can't let the threshold be too loose either. After all, we don't know which trajectory will end up winning at present time (we may be incurring more losses than necessary).

This separation of losers and winners - with drawdowns given available information - will form the basis of our methodology.

\section{Construction}

We now begin developing our framework for systematically finding stop-loss thresholds. We will need 2 components: the magnitude of the drawdowns themselves, and the categorization of that trades' trajectory as a winner or loser.

\subsection{Drawdown Distribution}

For each roundtrip trade executed by the Signal-Only strategy, we measure the maximum drawdown (as returns). Given the discrete nature of these results, we will bin the drawdowns into equal-size bins corresponding to their magnitude. Denote the set of these bins as $\{B_i\}$, with $i=1,\cdots,n$. This corresponds to the available information we will choose to analyze as quantitative researchers.

Note, however, that we could run into issues of sparsity. If a system trades very little, then we will be hard-pressed to find an $n$ such that we obtain meaningful results. This is perhaps the strongest limitation of our method.

\begin{figure}[h!]
	\centering\includegraphics[width=\columnwidth]{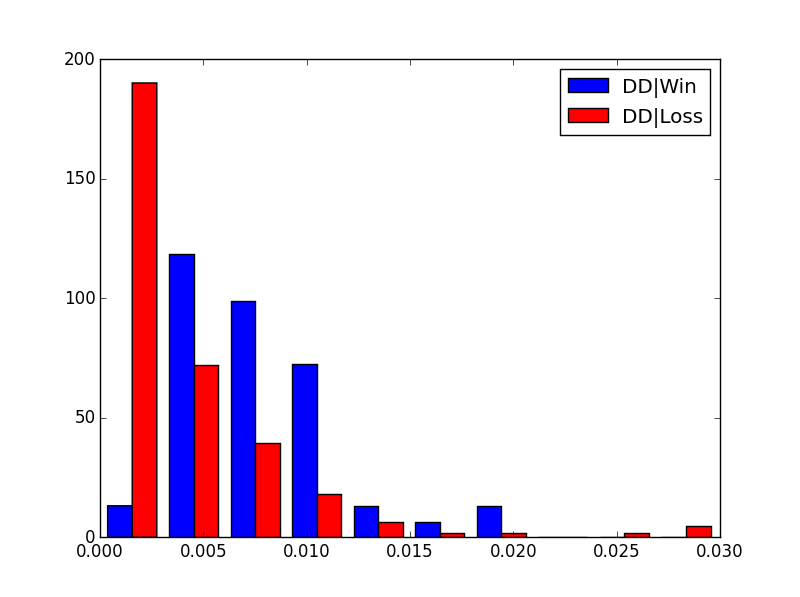}
	\caption{Conditional drawdowns for Signal-Only method on SPY.}\label{fig:hist}
\end{figure}

Plotting a histogram of the winning and losing drawdowns might be beneficial to illustrate the next steps - let's look at Figure~\ref{fig:hist}.
Say we enter a position in our live trading strategy and monitor it. We measure the drawdown to be $0.0025$. Intuitively, this means we would expect our position to behave like those in the 1st bin. As the drawdown changes, we can place our position in the appropriate bin and judge whether we believe that it will end up losing or winning based on the frequency of wins and losses from the Signal-Only measurements (in this example, we would expect a loss to be more likely).

However, if we set our stop threshold to be $0.01$, then our live positions could end up in any bin leading up to that threshold. In this example, we would incur losses more frequently until reaching the 2nd bin (the first where we are more likely to win). These losses are smaller, but it is important to account for all the bins leading up to our threshold.

In general, we wish to identify drawdown intervals for which a position is more likely to end up being a winner. We can formalize this argument by returning to the notion of expectation maximization.

\subsection{Return to Expected Values}

We now revisit our intuitive arguments and formalize them. Formally, the value we are looking to ascertain is the expected return ($r$) of a position given its (currently measured) maximum drawdown ($D$), or,
\begin{equation}
E\left( r | D\in B_i \right).
\end{equation}
We denote winning and losing positions by $W,L$ respectively and indicate our bins by $\{B_i\}$. Transforming our histogram into equations, we get
\begin{align}\label{eq:Er}
E\left( r | D\in B_i \right) =&{} E(r|L,D\in B_i) P(L|D\in B_i)\nonumber\\
&+ E(r|W,D\in B_i) P(W|D\in B_i),
\end{align}
where the RHS of equation~\eqref{eq:Er} is measured from the Signal-Only data.

As previously mentioned, when we enter a position, our drawdown can take on any value between $0$ and our stop threshold. As such, we need to try to incorporate the likelihood that a drawdown ends up in specific bins, and the returns we could expect in those cases:
\begin{equation}\label{eq:cumsum}
E(r|D\leq \ceil{B_k}) = \sum_{i=1}^{k}{E\left( r | D\in B_i \right) P(D\in B_i)},
\end{equation}
where $D\leq \ceil{B_k}$ signifies that $D$ is at most equal to the right endpoint of the bin $B_k$, and we once again measure from Signal-Only trades. Computationally, the expression in equation~\eqref{eq:cumsum} is analogous to computing the cumulative sum of the vector whose elements are the argument of the sum in~\eqref{eq:cumsum}. That is, we compute a vector of length $n$ whose $i$th element is $E\left( r | D\in B_i \right) P(D\in B_i)$. We then take the cumulative sum of that vector.

Since our overall objective is to maximize expected returns, we only need to take the threshold that corresponds to the highest value in the aforementioned vector. This gives us our stop threshold
\begin{equation}\label{eq:T}
 T = \argmax_{ \ceil{B_k}} \left\{ E(r|D\leq \ceil{B_k}) \right\}_{k=1}^n
\end{equation}

\section{Results}

We will now examine our method on market data from October 17, 2014 to July 22, 2016 - investing \$100,000.00 into the asset. We use the same Signal-Only strategy described in Section I, and we implement trailing stops calibrated to $T$ as defined in equation~\eqref{eq:T}, ie, exit the position if the price breaches $1-T$ times the maximum observed price since entry.

Note that it is not our intention to outperform the market, or a buy-and-hold strategy (which we include for reference), simply to ascertain if the systematic stop method outperforms the Signal-Only method.

\begin{figure}[h!]
	\centering\includegraphics[width=\columnwidth]{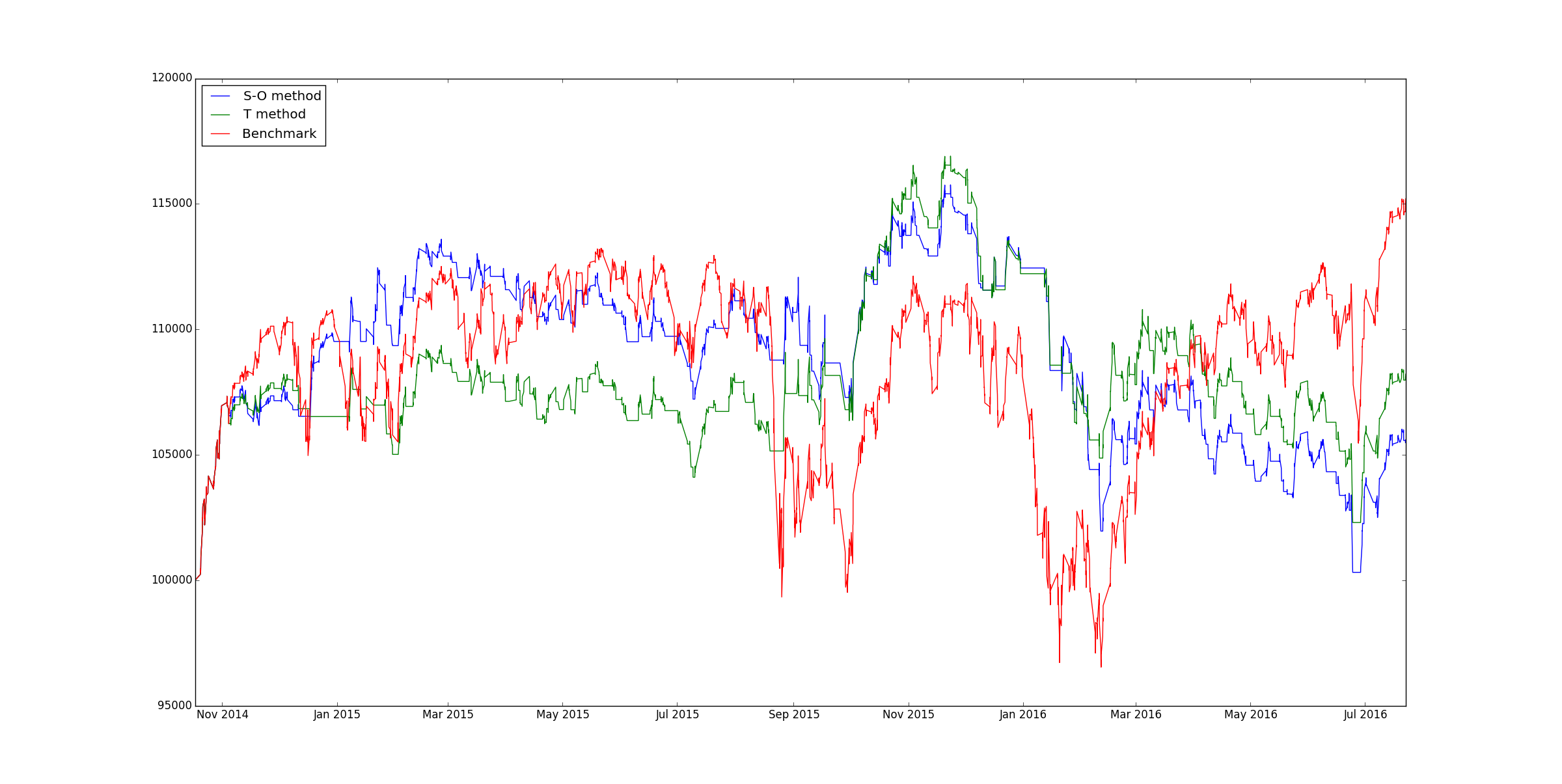}
	\caption{Results of $T$ method for SPY.}\label{fig:spy}
\end{figure}

No work in quantitative finance would be complete without an examination of performance on the S\&P 500 index, or in this case the SPY ETF. As we can see in Figure~\ref{fig:spy}, our stopping strategy outperforms the Signal-Only method. However, there are periods of under-performance.
We provide some additional results in Figure~\ref{fig:iwm}, and as a table in the Appendix.

\begin{figure}[h!]
	\centering\includegraphics[width=\columnwidth]{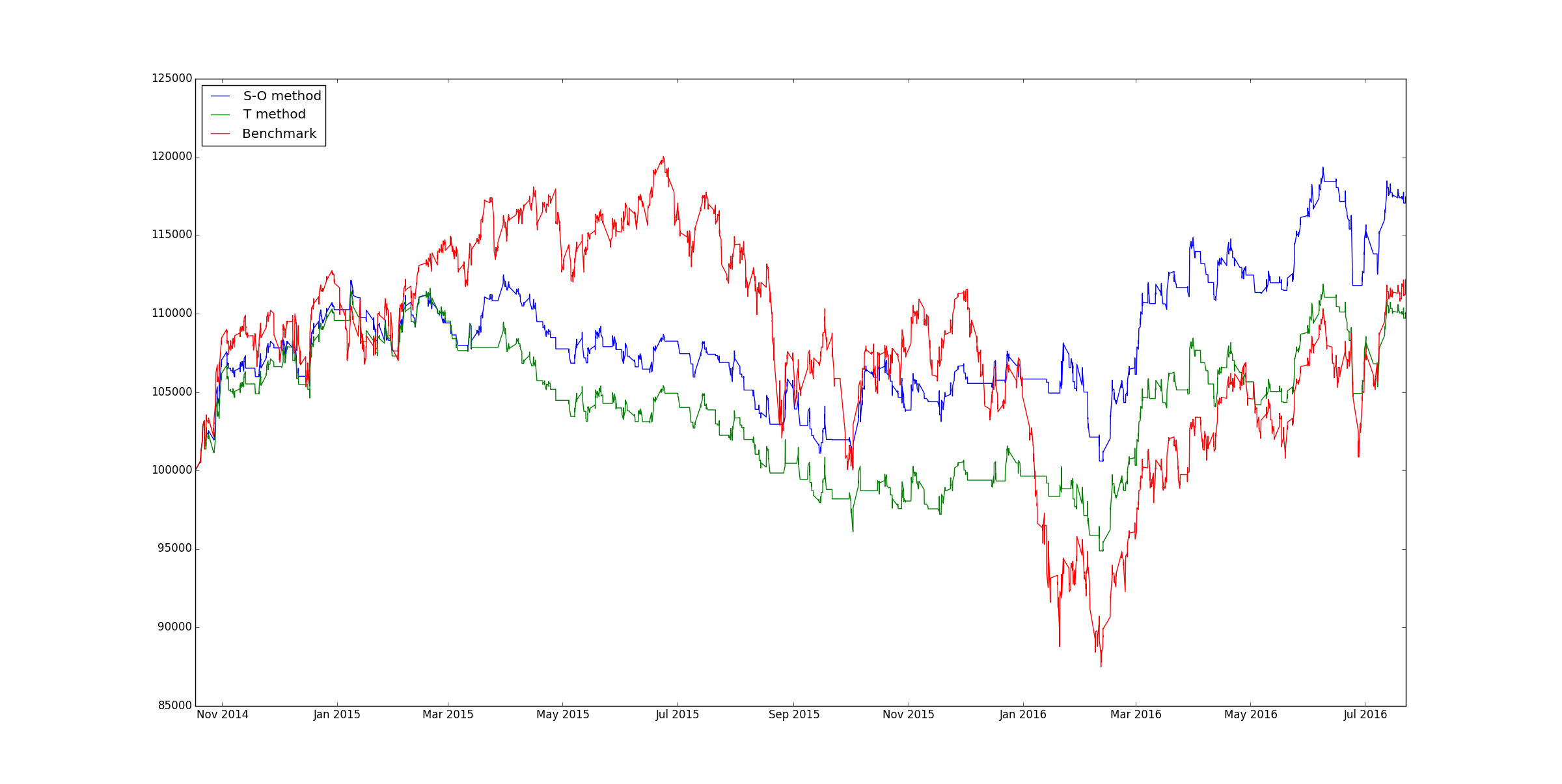}
	\caption{Results of $T$ method for IWM.}\label{fig:iwm}
\end{figure}

We immediately notice some troubling results. While we may or may not outperform the asset itself, we do not always outperform the Signal-Only strategy. In the Appendix, we present results for a mix of 114 liquid ETFs, large-cap, medium-cap and small-cap stocks. Our method outperforms the Signal-Only system in only $51.75\%$ of cases. In those cases, it will on average give rise to a $7.16\%$ increase in final net liquidation value (NLV). Unfortunately, losing cases provide an average $-8.20\%$ loss, resulting in an overall negative expected gain.

\subsection{Error Analysis}

Reflecting on our method, we recall that sparsity of data - that is Signal-Only systems that trade infrequently - could lead to spurious conclusions. Indeed, we may be binning very few values and drawing spurious conclusions about frequencies. In addition, too few measurements would also lead to larger standard errors in the measurement of expected returns.

We can check this by regressing the number of round-trip trades executed by our Signal-Only system against the change in final NLV (ie, the second and first columns of the Appendix). A scatter plot is shown in Figure~\ref{fig:corr}.
\begin{figure}[h!]
	\centering\includegraphics[width=\columnwidth]{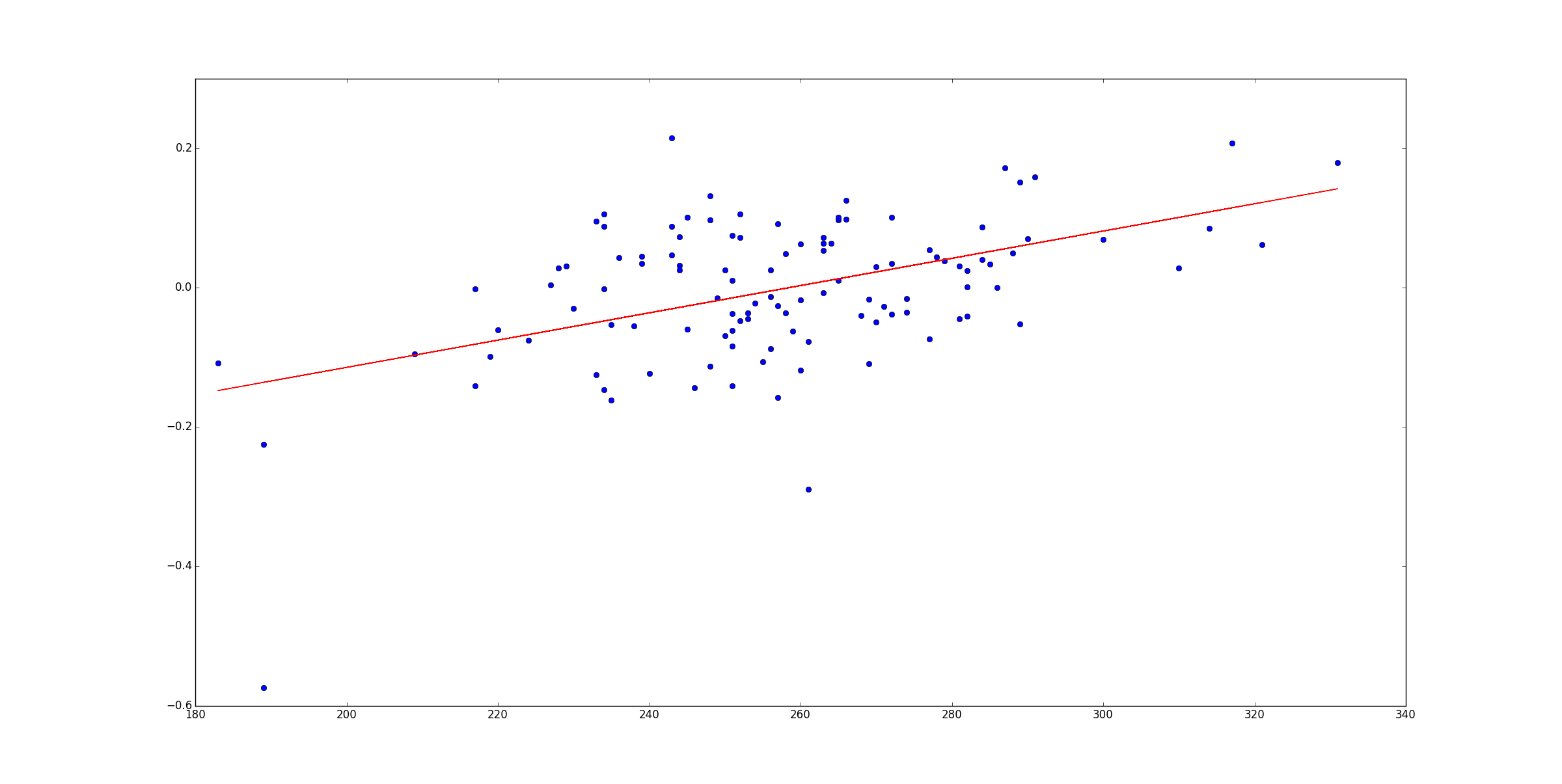}
	\caption{Change in NLV vs. number of trades.}\label{fig:corr}
\end{figure}

The statistics seem to support our hypothesis that there is a connection, with a Pearson correlation coefficient of $\rho = 0.4778$, and a $p$-value of $p=7.6\cdot 10^{-8}$ - which strongly rejects the null that the slope is $0$.

A more detailed study on the size of standard errors in our measurements is recommended for future work, as well as the calibration of $n$ to provide the best binning of our data once its size is determined. Study of this method acting on a particular subset of assets (ie, only large-cap) could also prove beneficial to practitioners.

\begin{figure}[h!]
	\centering\includegraphics[width=\columnwidth]{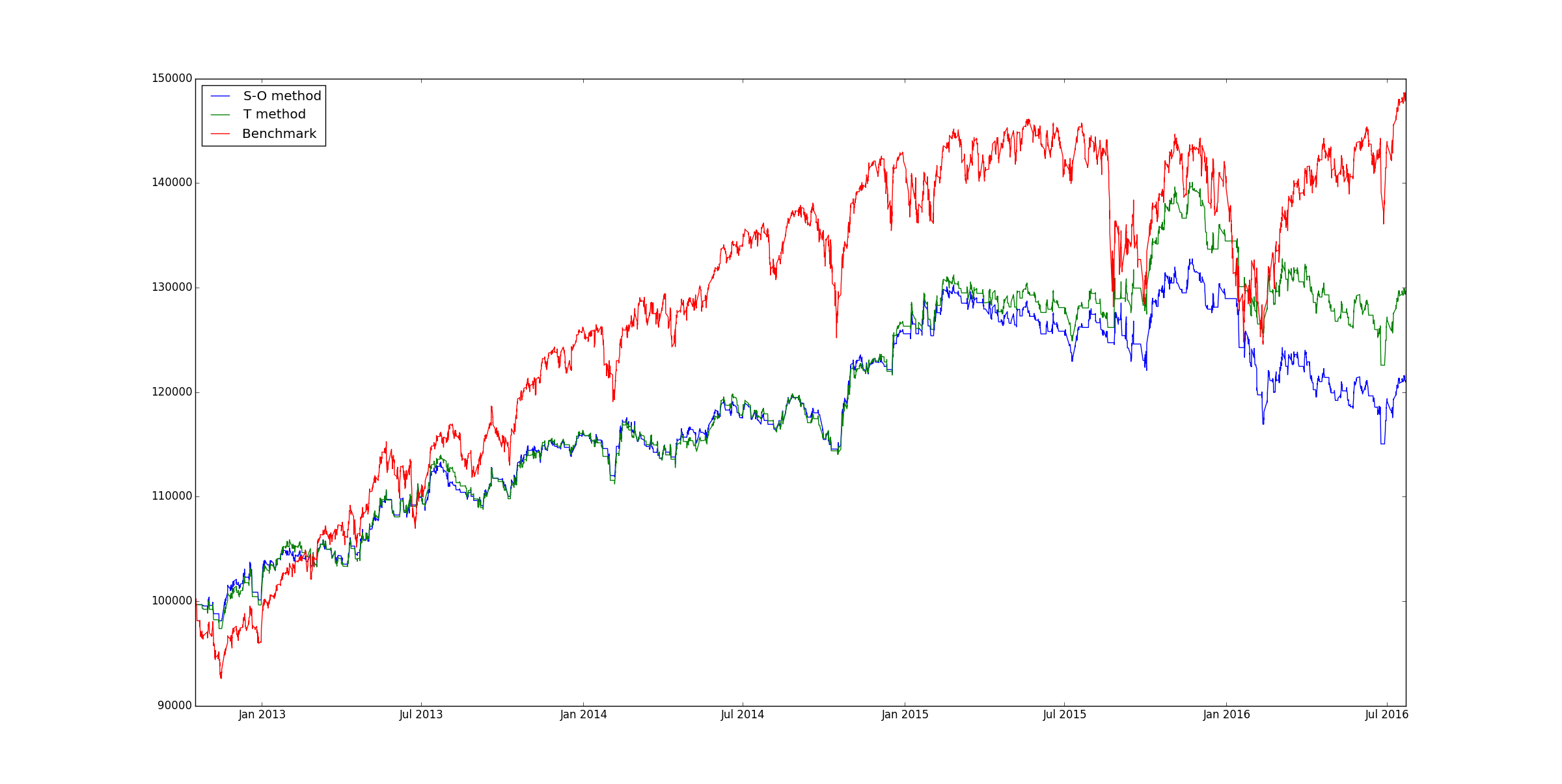}
	\caption{Long SPY.}\label{fig:spylong}
\end{figure}

In the meantime, we can quickly check whether more trade data is beneficial on SPY (for which we have more data), presented in Figure~\ref{fig:spylong}. At least on this singular data point, our method does seem to work with more data at its disposal. One possible shortfall is that going too far back in time might make the Signal-Only data ``stale", which is something to keep in mind. Since we don't have data reaching that far back for other assets, we will modify our method in the hopes of making it more flexible.

\subsection{Rolling Window}

In order to address the shortcomings of the original $T$ method, we need to effectively increase the density of our data. This can be done by focusing our analysis on the entry signal and the overall dynamics of the asset over a window of time.

Instead of looking at the Signal-Only trade points, we use every hourly point as a possible entry point and set the corresponding exit point to be $l$ hours ahead. So for each hourly point, we check if the asset price is above the 20-hour SMA, if so we enter a Long position which we exit in $l$ hours. We then take the last $m$ round-trip trades conducted this way to run our analysis and construct our threshold.

The threshold, call it $R$, is constructed as in equation~\eqref{eq:T}, the only difference is the data contained in $\{B_i\}$. In essence, we are analyzing the dynamics of the asset on a rolling window of time (in $l$-hour increments) and focusing on that, combined with our entry signal.

These ``trades" we are making to construct our stop threshold overlap with one another and do not account for any exit-signals. However, they provide a much larger number of data points to work with that will hopefully not be too stale. On the other hand, we now have to determine an appropriate $l$ and $m$. We set $l=20$ hours, a full SMA period, and set $m=250$ trades (somewhere near the center of our data from Figure~\ref{fig:corr}).

We provide results of this $R$ method in Figures~\ref{fig:spyroll} and~\ref{fig:iwmroll}, as well as in the Appendix. Comparing the $T$ and $R$ methods (Figure~\ref{fig:spy} vs~\ref{fig:spyroll} and~\ref{fig:iwm} vs~\ref{fig:iwmroll}). We see a clear improvement on both assets.

\begin{figure}[h!]
	\centering\includegraphics[width=\columnwidth]{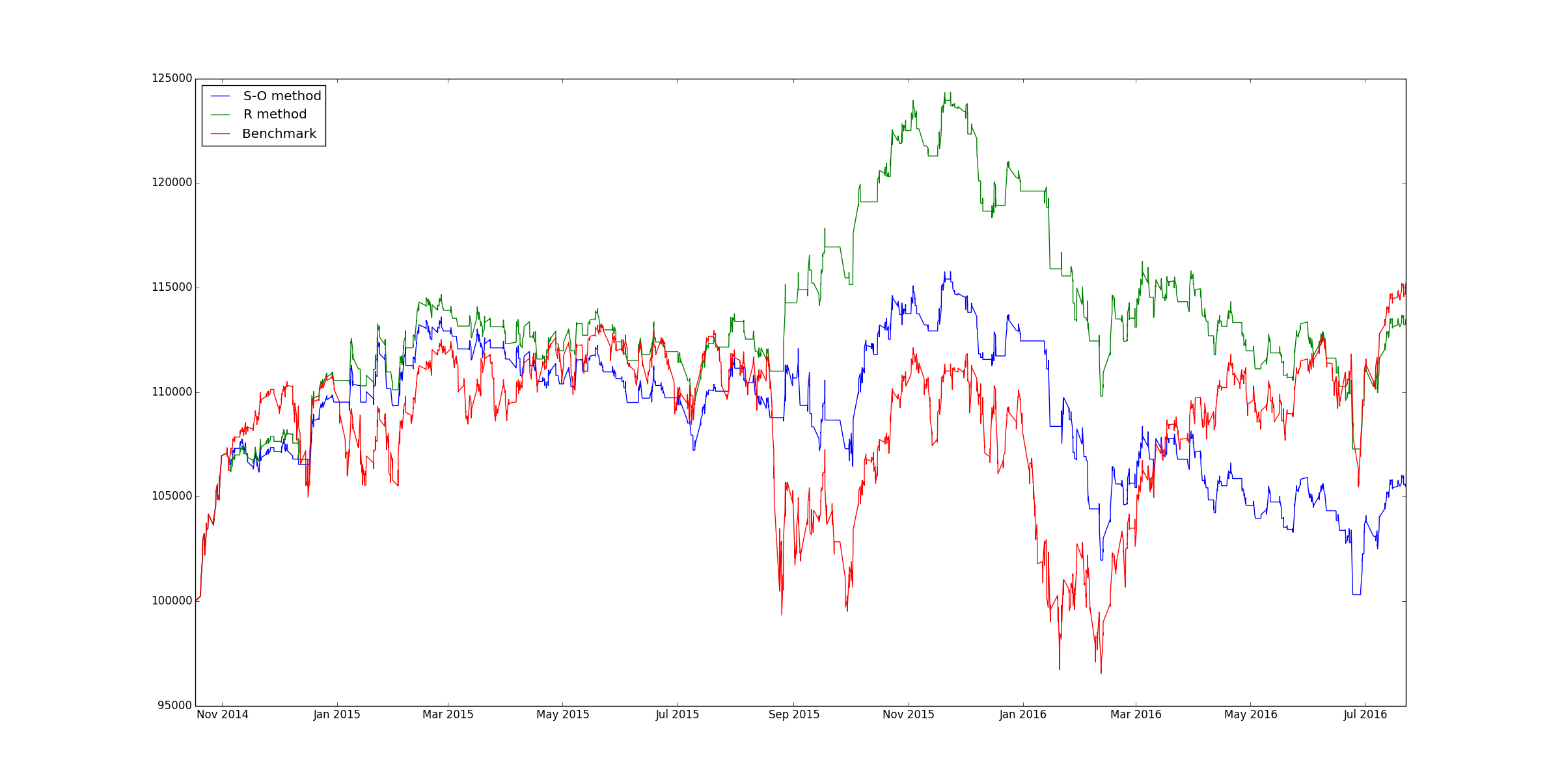}
	\caption{Results of $R$ method for SPY.}\label{fig:spyroll}
\end{figure}

\begin{figure}[h!]
	\centering\includegraphics[width=\columnwidth]{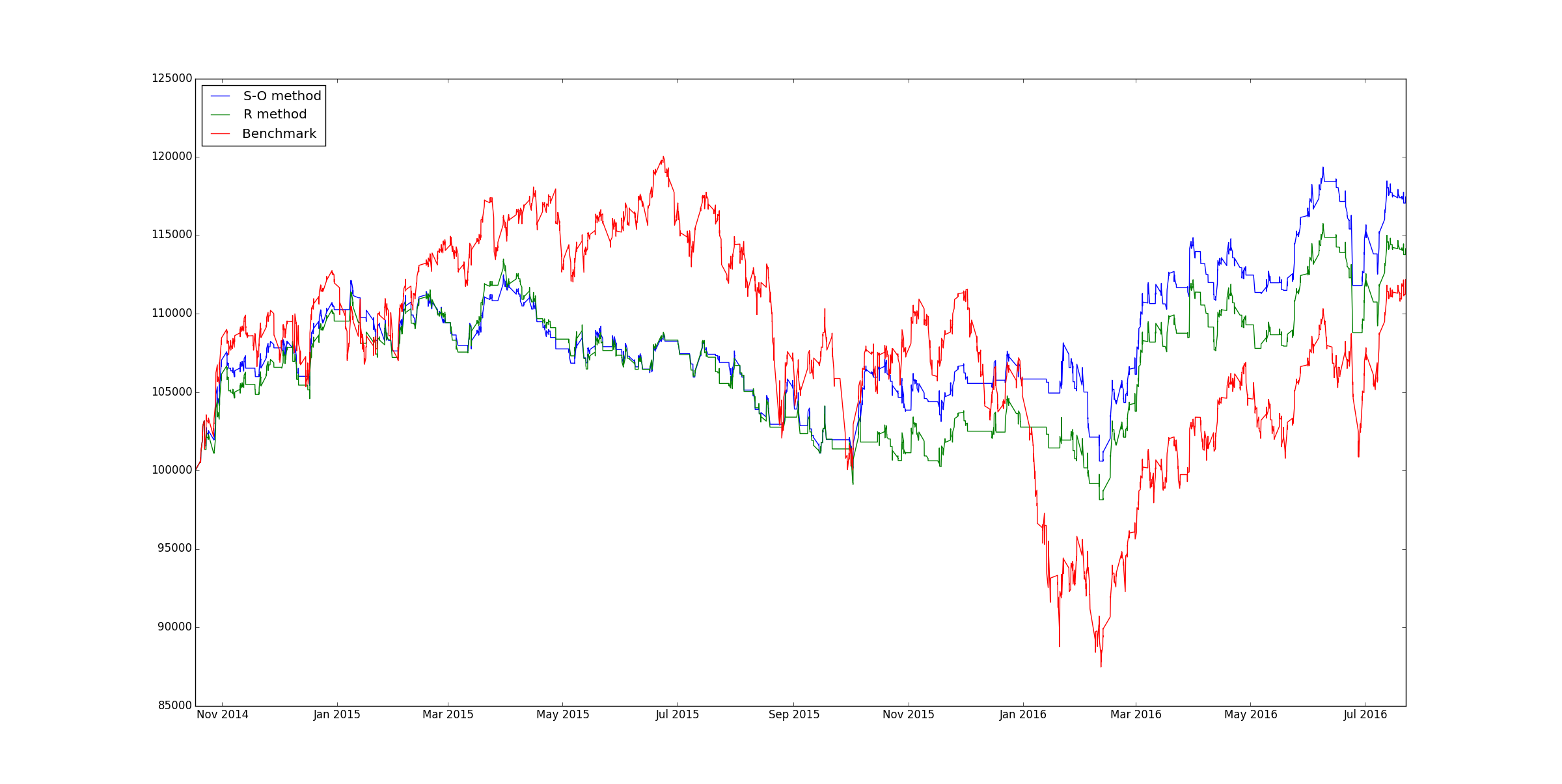}
	\caption{Results of $R$ method for IWM.}\label{fig:iwmroll}
\end{figure}

Looking again to our 114 names, we see an improvement across the board. We now see an increase in final NLV in $57.02\%$ of cases, with average increases of $6.37\%$ and average losses of $-6.94\%$. This gives us an overall expected change in NLV from Signal-Only to $R$ methods of $0.65\%$.

While our system is clearly not perfect, we believe the evidence of its merit is undeniable. As possible areas to explore to improve it further, we would recommend looking into calibrating $l$ to the mean or mode holding period for the Signal-Only method.

\subsection{Summary and Implementation}

We now summarize both methods and discuss some implementation aspects to facilitate extension. Firstly, the parameters:
\begin{enumerate}
	\item $n$, the number of bins. This can likely be chosen based on the number of trades (ie, square root method), though it is not yet known whether it should be the same value for all assets.
	\item $m$, the number of ``trades" to analyze in the $R$ method. This may also be varied based on the number of trades that yields good results (ie, Figure~\ref{fig:corr}).
	\item $l$, the holding period of ``trades" in the $R$ method. This can possibly be optimized based on the statistics of the holding periods in the Signal-Only method.
\end{enumerate}

Note that backtesting these methods can be computationally intensive, but that in live use, we would need only to update a few points at a time and it is unlikely to affect performance outside of a high-frequency trading environment.

\subsubsection{$T$ method}

The $T$ method uses trade data from the Signal-Only method. In essence, run a Signal-Only backtest up to the current time and store the maximum drawdown of each trade.  From that, we define bins $\{B_i\}$.

We then compute the probabilities of winning, losing and the respective average returns conditioned on being in each bin (equation~\eqref{eq:Er}). If these expected values are stored in a vector, then we construct $T$ by taking the cumulative sum of that vector, and getting the index of the maximum value. The right limit of the bin corresponding to that index is our $T$.

For any trade beyond that point (until we update), we set the stop threshold to be $1-T$ times the maximum observed price since entry.

\subsubsection{$R$ method}

The $R$ method uses artificial ``trades" and provides a greater number of data points. Using the data available until current time, take every potential trade entry point (in our case, hourly points) and check if the signal would identify an entry. If so, set the exit time of that ``trade" to be $l$ points ahead and record the drawdown.

Define bins $\{B_i\}$ and compute the necessary expected values as in the $T$ method. We set the stop threshold in the same way as well.

\section{Conclusion}

We have successfully constructed stop-loss thresholds in a systematic way and tested them on 114 assets. While a relatively small sample, this allows us to conclude that our method is on average quite successful, but imperfect.

It relies on the assumption that past trade data (real or artificial) will be indicative of the asset's behavior in the future. In a sense, we are saying that our signal will behave similarly to what it has done in the past.

Some components to examine in future work include the optimization of parameters $n,m,l$ and a close examination of those assets that did not perform well with our method. A careful look at how those compare to categories highlighted by~\cite{kam} is of particular interest.

While the $T$ method appears more intuitive, it suffers when using too little data. On the other hand, the $R$ method requires two additional parameters to be chosen. Both can perform well for some assets with an overall increase in performance when using the $R$ method - which reaches the critical point of a positive expected gain from its use. Lastly, given their systematic construction, this allows stop-loss thresholds to evolve with changing market conditions.

\ifCLASSOPTIONcaptionsoff
  \newpage
\fi



\bibliographystyle{IEEEtran}
%

\newpage
\appendix

We present all the data in two tables. In each, The first column corresponds to the percent change in net liquidation value at final time between the $T$ method and that Signal-Only method:
\begin{equation}
\Delta NLV_{TS} = \frac{NLV_T(t=t_f) - NLV_S(t=t_f)}{NLV_S(t=t_f)}.
\end{equation}
The second column is the percent change in net liquidation value at final time between the $R$ method and that Signal-Only method:
\begin{equation}
\Delta NLV_{RS} = \frac{NLV_R(t=t_f) - NLV_S(t=t_f)}{NLV_S(t=t_f)}.
\end{equation}
The third is the number of round-trip trades made by the Signal-Only system over the entire period.

\begin{table}[h!]
	\caption{First half of data.}
	\centering\begin{tabular}{lrrr}
		\toprule
		{} &  $\Delta NLV_{TS}$ & $\Delta NLV_{RS}$ &  Trades \\
		\midrule
		AAPL  & -0.099310 &      -0.057711 &         219 \\
		ADSK  & -0.161878 &      -0.170516 &         235 \\
		AEP   &  0.024071 &       0.043622 &         282 \\
		AMT   &  0.105779 &       0.014302 &         234 \\
		AXP   &  0.044818 &       0.074721 &         239 \\
		BA    & -0.001810 &       0.046409 &         234 \\
		BDX   &  0.033300 &       0.035577 &         285 \\
		BIG   &  0.097061 &       0.037862 &         265 \\
		BRK\_B &  0.100421 &       0.090277 &         272 \\
		CAH   &  0.063538 &       0.054561 &         263 \\
		CAT   &  0.095505 &       0.140478 &         233 \\
		CBS   & -0.052029 &      -0.031568 &         289 \\
		CCE   & -0.037289 &       0.043019 &         251 \\
		CI    & -0.157793 &      -0.113627 &         257 \\
		CMI   & -0.118370 &       0.024327 &         260 \\
		COP   & -0.113011 &      -0.198957 &         248 \\
		CSC   &  0.091598 &       0.047145 &         257 \\
		CTXS  &  0.031742 &       0.085372 &         244 \\
		CVX   & -0.084286 &      -0.146479 &         251 \\
		DE    &  0.100855 &       0.070411 &         265 \\
		DIA   &  0.025030 &       0.016271 &         244 \\
		DISCA & -0.027060 &       0.013972 &         271 \\
		DUK   &  0.073079 &       0.090376 &         244 \\
		ECL   & -0.017524 &       0.118206 &         260 \\
		EDC   & -0.574054 &      -0.095637 &         189 \\
		ED    & -0.041686 &      -0.006473 &         282 \\
		EL    & -0.069665 &      -0.074991 &         250 \\
		EMN   & -0.026522 &       0.030367 &         257 \\
		EMR   & -0.044531 &       0.020131 &         253 \\
		EQR   & -0.038693 &      -0.036966 &         272 \\
		ETN   & -0.061544 &      -0.095397 &         251 \\
		EWW   & -0.108622 &      -0.118523 &         183 \\
		EWZ   & -0.225347 &      -0.049079 &         189 \\
		FB    &  0.034308 &       0.067675 &         272 \\
		FDX   &  0.028010 &       0.031280 &         228 \\
		FE    & -0.045101 &      -0.034312 &         281 \\
		FLS   & -0.095633 &      -0.056059 &         209 \\
		FL    & -0.017038 &       0.047850 &         269 \\
		GOOG  & -0.088212 &       0.022924 &         256 \\
		HCN   & -0.000551 &      -0.102810 &         286 \\
		HD    & -0.049910 &      -0.064931 &         270 \\
		HON   & -0.047627 &       0.030111 &         252 \\
		HOT   &  0.043231 &      -0.079104 &         236 \\
		HUM   &  0.084557 &      -0.033134 &         314 \\
		IBM   & -0.054980 &      -0.012029 &         238 \\
		IJH   & -0.002449 &      -0.004025 &         217 \\
		INTU  &  0.171880 &       0.106334 &         287 \\
		ITW   & -0.030335 &      -0.003375 &         230 \\
		IVV   &  0.010142 &       0.083939 &         251 \\
		IWD   & -0.014885 &       0.031965 &         249 \\
		IWM   & -0.062709 &      -0.028134 &         259 \\
		IWN   & -0.036429 &       0.043210 &         258 \\
		IWO   & -0.143538 &      -0.107051 &         246 \\
		JCI   &  0.087559 &       0.059123 &         243 \\
		JNJ   &  0.039887 &      -0.018860 &         284 \\
		KBE   &  0.105291 &       0.082846 &         252 \\
		KLAC  &  0.096844 &       0.027069 &         248 \\
		\bottomrule
	\end{tabular}
\end{table}
\begin{table}[h!]
	\caption{Second half of data.}
	\centering\begin{tabular}{lrrr}
		\toprule
		{} &  $\Delta NLV_{TS}$ & $\Delta NLV_{RS}$ &  Trades \\
		\midrule
		KMB  &  0.070250 &       0.042439 &         290 \\
		KRE  &  0.063593 &       0.043243 &         264 \\
		KR   &  0.034545 &       0.014848 &         239 \\
		KSS  & -0.123110 &      -0.016322 &         240 \\
		LEG  &  0.074460 &      -0.019758 &         251 \\
		LLY  & -0.035339 &       0.091602 &         274 \\
		LMT  &  0.061248 &       0.067681 &         321 \\
		LOW  & -0.059598 &      -0.137631 &         245 \\
		LVS  & -0.061214 &      -0.091064 &         220 \\
		MAC  &  0.053715 &      -0.047753 &         277 \\
		MA   &  0.062743 &       0.055563 &         260 \\
		MCO  &  0.043971 &       0.088168 &         278 \\
		MDY  &  0.003640 &       0.008749 &         227 \\
		MJN  &  0.150893 &       0.053198 &         289 \\
		MRK  &  0.207356 &       0.132181 &         317 \\
		MSI  &  0.038377 &       0.035478 &         279 \\
		MTW  &  0.029345 &      -0.012442 &         270 \\
		MUR  & -0.289242 &      -0.110098 &         261 \\
		MYL  &  0.097993 &       0.052649 &         266 \\
		NEE  & -0.039901 &       0.004521 &         268 \\
		NOC  & -0.036898 &      -0.038880 &         253 \\
		PCAR & -0.125426 &      -0.132029 &         233 \\
		PH   & -0.022884 &      -0.068078 &         254 \\
		PNC  &  0.100277 &       0.048727 &         245 \\
		PNR  & -0.075399 &      -0.133893 &         224 \\
		PRU  & -0.078001 &      -0.105371 &         261 \\
		PX   &  0.046591 &       0.023065 &         243 \\
		QCOM &  0.087612 &       0.047637 &         234 \\
		QQQ  & -0.013143 &      -0.010100 &         256 \\
		ROK  &  0.071564 &       0.050945 &         263 \\
		ROST &  0.179216 &       0.105176 &         331 \\
		RTN  &  0.027442 &       0.048043 &         310 \\
		SBUX &  0.086377 &       0.141163 &         284 \\
		SEAS & -0.141261 &      -0.160344 &         251 \\
		SMH  &  0.030827 &      -0.013101 &         229 \\
		SPY  &  0.024745 &       0.074618 &         256 \\
		SSO  &  0.131903 &       0.274300 &         248 \\
		STI  &  0.071474 &       0.067970 &         252 \\
		STZ  & -0.106407 &      -0.098256 &         255 \\
		TAP  & -0.109367 &      -0.122712 &         269 \\
		THC  &  0.214121 &       0.169901 &         243 \\
		TMO  &  0.068488 &       0.086475 &         300 \\
		TWX  &  0.125078 &       0.074304 &         266 \\
		UNM  &  0.158868 &       0.132328 &         291 \\
		UNP  & -0.053335 &      -0.052079 &         235 \\
		UPS  &  0.031115 &      -0.019336 &         281 \\
		URBN &  0.001012 &       0.008519 &         282 \\
		UTX  & -0.016383 &       0.064200 &         274 \\
		VLO  &  0.049661 &       0.088453 &         288 \\
		VNQ  & -0.007583 &       0.027552 &         263 \\
		VOO  &  0.048696 &       0.131937 &         258 \\
		VTR  & -0.073986 &      -0.026931 &         277 \\
		V    &  0.053293 &       0.053392 &         263 \\
		WDC  & -0.141532 &      -0.096587 &         217 \\
		WEC  &  0.009663 &      -0.031661 &         265 \\
		XLE  &  0.024969 &      -0.104159 &         250 \\
		XRT  & -0.147019 &      -0.010695 &         234 \\
		\bottomrule
	\end{tabular}
\end{table}

%
\newpage






\end{document}